**Dust Stratification in Young Circumstellar Disks**

Terrence Rettig[1], Sean Brittain[2], Theodore Simon[3], Erika Gibb[4], Dinshaw S. Balsara[1], David A. Tilley[1], Craig Kulesa[5]

[1] Center For Astrophysics, University of Notre Dame, Notre Dame, IN 46556 (trettig@nd.edu)
[2] NOAO, Tucson, AZ 85719
[3] Institute for Astronomy, University of Hawaii, Honolulu, HI 96822
[4] Department of Physics and Astronomy, University of Missouri-St. Louis, St. Louis, MO 63121
[5] Department of Astronomy, University of Arizona, Tucson, AZ 85723

ABSTRACT - We present high-resolution infrared spectra of four YSOs (T Tau N, T Tau S, RNO 91, and HL Tau). The spectra exhibit narrow absorption lines of $^{12}CO$, $^{13}CO$, and $C^{18}O$ as well as broad emission lines of gas phase $^{12}CO$. The narrow absorption lines of CO are shown to originate from the colder circumstellar gas. We find that the line of sight gas column densities resulting from the CO absorption lines are much higher than expected for the measured extinction for each source and suggest the gas to dust ratio is measuring the dust settling and/or grain coagulation in these extended disks. We provide a model of turbulence, dust settling and grain growth to explain the results.

Subject headings: T Tauri stars, circumstellar matter --- stars: formation --- stars: individual (HL Tau, RNO 91, T Tau S, T Tau N)
stars: pre-main-sequence --- protoplanetary disks

**1. Introduction**

The physical structure of extended flared disks around YSOs (young stellar objects) depends on a variety of physical processes, including the extent to which the gas and dust are mixed by turbulence, the rate at which the dust is able to separate from the gas and settle to the disk mid-plane, and the effectiveness of grain coagulation to build grains of increasing size while avoiding fragmentation. Although it is difficult to explore all of them observationally, these topics have drawn a substantial amount of theoretical interest, in particular with regard to the importance of dust sedimentation and mid-plane turbulence to the formation of planetesimals (Goldreich & Ward 1973; Safronov 1969; Weidenschilling 1995; Goodman & Pindor 2000; Youdin & Shu 2002; & Youdin & Chiang 2004).

In general, dust settling and grain growth are expected to occur throughout the upper disk atmosphere (e.g., Miyake & Nakagawa 1995; D'Alessio et al. 1998, 1999, 2001; Chiang et al. 2001; Dullemond and Dominik 2004) until a balance is reached with diffusion (turbulence) near the mid-plane. In a recent model, Dullemond and Dominik (2005) incorporate grain coagulation and settling in a protoplanetary disk to compare with the dust images. They find that the coagulation of small particles occurs much too fast to be consistent with disk ages and observed SEDs and suggest grain fragmentation is needed to replenish the dust. Whether fast or slow, the general effect of grain growth will be to decrease the line of sight column density of particles (i.e., extinction) at high disk latitudes and increase the density near the mid-plane. Thus, the question becomes, as the particles coagulate and settle to the midplane what happens to the line of sight gas/dust ratio? Is the turbulent mixing in the midplane sufficient to maintain a constant gas to dust ratio over some or all of inclinations or will the depletion of dust be faster than gas depletion?

Observations and models of protoplanetary disks are increasingly successful at accounting for many of the observed properties but observational constraints have remained elusive for the overall structure of gas relative to the dust. The temperature and mass of the inner warm gas (R<1AU) around a number of YSOs has been derived from infrared observations of molecules (Carr et al. 2001; Blake and Boogert 2004; Najita et al. 2003, Brittain et al. 2003, Rettig et al. 2004), the distribution of the cold gas in the outer regions (R>100AU) from mm and sub-mm observations (e.g., Sargent and Beckwith 1994), and the distribution of dust in flared disks from spectral energy distributions as well as optical/infrared imaging (Kenyon and Hartman 1987, Chiang & Goldreich 1997, 1999; Chiang et al. 2001; D'Alessio et al. 1999, 2001, 2005). The sizes of dust grains in circumstellar disks have been constrained by infrared to sub-mm to cm observations (Van Boekel et al. 2004; Mannings and Sargent 1997; Rodmann et al. 2005; Wilner et al. 2005, Furlan et al. 2005). Optical/IR imaging for GG Tau has provided some evidence for grain growth and/or stratification and constraints on dust sizes in a very limited vertical region above the midplane (Duchene et al. 2004). Similarly, 11.8 µm images of HK Tau B indicate the presence of a significant reservoir of 1.5 – 3.0 µm grains at the surface of the disk (McCabe, Duchene, and Ghez 2003). Evidence of dust settling has been suggested by Glassgold et al. (2004) where they suggest that since the $H_2$ in the disk of TW Hya is UV pumped, the dust must have settled.

In this paper, we present direct observational comparisons of gas and dust in disks around YSOs. Since we cannot observe a single source at different inclinations, we use four young stellar objects (YSOs) (T Tau N, T Tau S, RNO 91, and HL Tau) having a variety of inclinations. Brittain et al. (2005) presented the first data for HL Tau that demonstrated the line of sight was gas rich, and suggested that dust settling may be responsible. The column of gas through the extended disk atmosphere is obtained by using high-resolution molecular line spectroscopy of CO. The dominant gas component of circumstellar disks is $H_2$, however, the ro-vibrational lines of $H_2$, which exhibit a weak quadrupole spectrum with steep excitation requirements, are very difficult to detect (Bary et al. 2003). In contrast, the ro-vibrational lines of CO have been shown to have excellent diagnostic characteristics over a wide range of temperatures (Brittain et al. 2003, Najita et al. 2003). Using the CO column density with measured inclinations and extinctions from previous work, we obtain line of sight gas/dust ratios as a function of inclination. The gas/dust ratio is reflective of only the smaller grains along our line of sight as the larger grains, essentially the refractory compounds and ices that are bound up in larger bodies have effectively been removed from the system. The resulting gas/dust ratios can be used to quantify the rate of disk stratification and provide initial constraints for theoretical models. The four sources considered in this work have similar ages and SEDs, and show active accretion, but are not in outburst. Section 2 presents high-resolution spectral observations of the four sources in this study. In Section 3 we present the gas/dust ratios for each source as a function of inclination. In Section 4 we present initial gas and dust models with turbulence and grain opacities that provide line of sight predictions for gas/dust ratios as a function of inclination. A comparison with the observations indicates both dust stratification and grain growth is needed. Potential implications for constraining midplane density and turbulence as well as consequences for planet formation models are discussed. We are aware of the obvious limitations for the inclination and extinction measures and stress the importance of a larger data sample.

## 2. Observations and Data Analysis

To determine the column density of gas for each source, we used the NIRSPEC spectrometer (McLean et al. 1998) at the W. M. Keck Observatory on Mauna Kea, Hawaii, to observe the near-infrared lines of CO at high dispersion (RP ~ 25,000). The analysis of the absorption lines of CO was completed in a manner similar to that described by Brittain et al. (2003, 2005). Flats and darks were used to remove systematic effects at each grating setting and the spectra were cleaned of systematically hot and dead pixels as well as cosmic ray hits. Absolute flux calibration was achieved through observations of standard stars. Table 1 provides a summary of the M- and K-band observation, spectral coverage, observed radial velocity, spectral setting and integration time for each of the sources.

**Table 1. Observations**

M-band

| Object | Date | Spectral range (cm$^{-1}$) | Vr km/s | Setting/Order | Integration Time (min) |
|---|---|---|---|---|---|
| T Tau N | 3/23/2002 | 2125-2158 | 47±1 | MW1/16 | 8 |
| HL Tau | 3/23/2002 | 2125-2158 | 48±1 | MW1/16 | 4 |
| RNO 91 | 3/18/2003 | 2125-2158 | -43±1 | MW1/16 | 12 |
| T Tau S | 3/23/2002 | 2125-2158 | 47±1 | MW1/16 | 8 |

K-band

| Object | Date | Spectral range (cm$^{-1}$) | Vr | Setting/Order | Integration Time (min) |
|---|---|---|---|---|---|
| T Tau N | 3/23/2002 | 4220-4245 | 47±1 | K/32 | 4 |
| HL Tau | 3/23/2002 | 4220-4245 | 49±1 | K/32 | 8 |
| RNO 91 | 7/31/2004 | 4220-4245 | 13±2 | K/32 | 16 |
| T Tau S | 3/23/2002 | 4220-4245 | 47±1 | K/32 | 4 |

In order to cancel most of the atmospheric and background effects due to night sky emission lines and the thermal (~300 K) continuum that dominates the M-band, we nodded the telescope by a small distance along the slit between two positions, A and B, in an {A, B, B, A} sequence. Combining the scans as (A-B-B+A)/2 canceled the telluric features to first order. Subsequently, the FWHM of the spatial profiles in the "A" and "B" rows were extracted to obtain the spectra for both positions. We used the Spectrum Synthesis Program (SSP, Kunde & Maguire 1974) and the HITRAN 2000 molecular database (Rothman et al. 2003) to model the atmospheric transmittance function of the combined spectrum. The optimized model established the column burden of absorbing atmospheric species, the spectral resolving power, and the wavelength calibration. The atmospheric model was divided into the spectra to reveal the stellar spectra for each source (Figure 1a,b). T Tau N and T Tau S are separated by ~ 0.7" on the sky so the extracted continuum and model psf is provided in Figure 2 to illustrate our ability to separate the two sources.

Three of the sources (T Tau S, RNO91, & HL Tau) have very similar SEDs and therefore are likely to be in approximately the same evolutionary state. T Tau N has the SED of a slightly more evolved YSO but is the visibly brighter companion to T Tau S and so presumably has very nearly the same age. The slightly different SED for T Tau N might be the result of its nearly face-on perspective.

The infrared spectra of HL Tau, RNO 91 and T Tau N (see Figure 1a,b) show broad CO ro-vibrational emission lines upon which are superposed a set of narrow absorption lines. The broad emission lines are evidence for very hot gas in the inner disk, which originates from a region that extends radially outwards from the co-rotation radius of the disk to ~1 AU (see Najita et al. 2003). The observation of these broad emission lines strongly suggests that the central hot disk region was being observed at infrared wavelengths. That the absorption lines have the same Doppler shift as the emission features to within a few km/sec (our calibration uncertainty) suggests the observed gas originates in an extended disk atmosphere (or envelope) and not from interstellar material.

The narrow low-J absorption lines of $^{12}CO$, $^{13}CO$, $C^{18}O$ near 4.7 μm and the overtone lines (v=2-0) at 2.3 μm are of particular interest since they provide a measure of the amount of cooler material along the line of sight to the star or inner disk. Using an excitation diagram for each source, we determined the rotational temperatures of the CO absorption lines for the four sources. The column density of $^{12}CO$ is typically derived from the optically thin overtone absorption lines at 2.3 μm (see Brittain et al. 2005).

A measure of the amount of dust along the line of sight to the central star is provided by the extinction, $A_V$. For the sources observed in this work, we make use of $A_V$ estimates that have been inferred from optical and broad-band near-infrared (JHK) photometry. Thus the infrared continuum against which the N(CO) is measured and against which the extinction has been inferred is identical. Direct comparison of the dust and gas content is therefore possible from the very same "pencil beam" absorbing column. Because dust particles are expected to settle to the mid-plane around YSOs, our observations of the gas/dust ratios for disks that are oriented at several different inclinations to the line of sight offer a direct measure of the extent to which the dust has settled to the mid-plane and they indirectly provide a constraint on the mid-plane turbulence.

Potentially some CO absorption could result from interstellar foreground material. However, in general the correction due to interstellar extinction will be small compared to the dust-to-gas variability that is expected as a function of inclination. Also, the CO excitation temperature and its Doppler shift can generally be used to determine whether the gas originates from the disk atmosphere, cold foreground gas (Rettig et al. 2005), or the stellar photosphere.

## 3. Underlying Scenario and Results

If the gas and dust in YSO disks is largely mixed (D'Alessio et al. 1999), then one might expect that $[N(CO)/A_V]_{disk}$ will be independent of the viewing angle. Alternatively, if dust preferentially settles to the mid-plane of such disks, there should be a correlation between $N(CO)/A_V$ and the inclination of the disk. Observations along a dust-depleted line of sight towards a face-on disk would be expected to show a larger $N(CO)/A_V$ ratio than would a disk that is viewed close to the mid-plane, whose ratio should approach the interstellar value. A comparison interstellar ratio of $[N(CO)/A_V]_{interstellar}$ is obtained from the measured $CO/H_2$ abundance ratio of $1.56 \times 10^{-4}$ and the $A_V/N(H_2) = 10.8 \times 10^{-22}$ mag cm$^{-2}$ (Mathis 1990). Kulesa (2002) has confirmed this relationship for $A_V/N_H$ in dense clouds with $A_v$ up to 60. Thus, an interstellar CO gas/dust ratio can be given as $[N(CO)/A_V]_{interstellar} = 1.4 \times 10^{17}$ cm$^{-2}$ mag$^{-1}$ which can be used to normalize the observed line of sight measurements of $[N(CO)/A_V]_{disk}$ for the sources with differing inclinations. The normalized value will be defined as $\Delta = [N(CO)/A_V]_{disk} / [N(CO)/A_V]_{interstellar}$ a measure of how much the line of sight differs from the interstellar value.

In the present study, we consider a sample of four YSOs. The parameters and results for each source are summarized below and in Table 2:

(1) T Tau N has a nearly face-on circumstellar disk, i.e., $i \sim 19°$ (Herbst, Robberto, & Beckwith 1997); $i \sim 23°$ (Eisloffel & Mundt 1998); and perhaps as much as $i \sim 45°$ (Stapelfeldt et al. 1998). The extinction, from optical and infrared measurements, is approximately $1.5 \pm 0.2$ mag (Cohen & Kuhi 1979; White & Ghez 2001). The nearly face-on projection T Tau N and its minimal extinction suggest only trace CO gas should be detected. However, the measured CO column density for this line of sight, $N(CO) = 1.7 \pm 0.1 \times 10^{18}$ cm$^{-2}$, is much larger than expected. With an excitation temperature of $\sim 100$ K, the gas is almost certainly associated with the flared disk. In general, gas at this temperature cannot be attributed to stellar absorption in the photosphere or cold foreground material but it should be noted that Martin et al. (2005) have shown the possibility that cloud turbulence can produce warm interstellar gas. The resulting $[N(CO)/A_V]_{disk} = 1.2 \pm 0.2 \times 10^{18}$ cm$^{-2}$ mag$^{-1}$ for T Tau N, so $\Delta = 8.1 \pm 1.5$ times the normal interstellar gas/dust ratio. Clearly, if the gas/dust ratio throughout the circumstellar environment began initially with an ISM value then this result suggests that dust settling and/or grain growth has occurred.

(2) RNO 91 is inclined by $\sim 60°$ (Weintraub et al. 1994) and has $A_V \sim 9$ (Myer et al. 1987. We measure the $N(CO) = 3.3 \pm 0.2 \times 10^{18}$ cm$^{-2}$, which includes a contribution from CO ice of $N(CO) \sim 2.8 \times 10^{17}$ cm$^{-2}$. Thus, we find $[N(CO)/A_V]_{disk} = 3.8 \pm 0.7 \times 10^{17}$ cm$^{-2}$ mag$^{-1}$, which is approximately 2.7 times the interstellar value. Where error estimates for $A_V$ are not included in the literature, we assume a 10-15% error, see Table 2.

(3) Similarly, HL Tau is inclined by $\sim 67°$ (Lucas et al. 2004), and it lies behind $\sim 24$ magnitudes of extinction (Close et al. 1997). We find $N(CO) = 7.5 \pm 0.2 \times 10^{18}$ cm$^{-2}$ (Brittain et al. 2005), thus $[N(CO)/A_V]_{disk} = 3.2 \pm 0.5 \times 10^{17}$ cm$^{-2}$ mag$^{-1}$. The resulting $\Delta$ for HL Tau is more than twice the interstellar value.

(4) T Tau S presents a nearly edge-on disk at $i > 80°$ (Solf & Bohm 1999; Walter et al. 2003) and a visual extinction of $A_v \sim 35$ mags (Koresko et al. 1997). Beck et al. (2004) suggested somewhat smaller extinction for T Tau S but the relationship between $A_V$ and the ice and silicate features is not well established for circumstellar material that is warmer and likely has ongoing grain processing. T Tau N has a nearly face-on circumstellar disk with a radius of $\sim 40$ AU (Akeson et al. 1998). Since the projected separation between the binary components of T Tau is $\sim 100$ AU (Hogerheijde et al. 1997; Akeson et al. 1998; Beck et al. 2004), the disk of T Tau N cannot be a source of extinction for T Tau S. Moreover, since the ages ($\sim 1$ million years) and distances of the two components are similar (Cohen & Kuhi 1979, Hartmann 1999), the different inclinations of the two disks are most likely responsible for the differences in extinction. T Tau S has been found to be a complicated system (Solf & Bohm 1999, Duchene et al. 2002, 2005) and large extinction variations have been noted (see Ghez et al. 1991, Koresko et al. 1997, Beck et al. 2004) so its extinction is probably the least well known in this data sample. For T Tau S, we measure N(CO)=9.0±0.3x10$^{18}$ cm$^{-2}$ (see also Duchene et al. 2005). We find [N(CO)/A$_V$]$_{disk}$ = 2.3±0.4x10$^{17}$ cm$^{-2}$ mag$^{-1}$ and thus Δ = 1.6±0.3, which is approaching the canonical interstellar gas-to-dust ratio.

**Table 2.**

| Star | Inclination | Disk N(CO)/A$_V$ cm$^{-2}$ mag$^{-1}$ | T(CO) | Δ |
|---|---|---|---|---|
| T Tau N | ~20-45° | 1.2±0.2x10$^{18}$ | ~100 K | 8.1±1.5 |
| RNO 91 | ~60±10° | 3.8±0.7x10$^{17}$ | ~50 K | 2.7±0.5 |
| HL Tau | ~67±10° | 3.2±0.5x10$^{17}$ | ~100 K | 2.3±0.4 |
| T Tau S | >80° | 2.6±0.4x10$^{17}$ | 100-300 K | 1.8±0.4 |

A$_V$ provides a measure of the extinction due to grains smaller than about 5μm. Error bars for the extinction for RNO 91, HL Tau, and T Tau S are not reported in the literature and are assigned 10-15% errors in the above calculations. Also, for inclinations, we include an estimated error of ±10° for RNO 91, HL Tau, and T Tau S as they are not reported in the literature.

Clearly, the trend in Δ shows that when the disk is viewed more face-on, as shown in Figure 3, the gas/dust ratio is higher and, conversely, as the line of sight approaches the mid-plane (e.g., T Tau S), Δ approaches the cosmic gas-to-dust ratio. Heavily embedded sources, such as T Tau S, might be more affected by scattered light so we might be underestimating the extinction. However, any scattering correction for the edge-on sources will only cause the trend to be more pronounced. We also have to note that the four sources might also be in different evolutionary states (the transition time from sources that are accreting from their envelope to their disk to Class II sources is very short, ~100,000 yrs). Whether this will affect the observed gas/dust ratio is not known but a larger sample of sources will help to clarify this issue. As a relevant comparison, HL Tau and RNO 91 have similar inclinations and similar gas/dust ratios, although the N(CO) and A$_V$ are 2-3 times lower for RNO 91. Whether the lower N(CO) and A$_V$ for RNO 91 are attributable to a minimal age difference or different initial conditions in the disk is unknown. But it does present an intriguing hint that the observed gas-to-dust ratio remains relatively constant for a given inclination during this transitional phase.

In principle, the anomalous gas/extinction ratios observed among these four stars could be due to: 1) settling of grains leading to a gas rich disk atmosphere, 2) growth of grains leading to a dust depleted line of sight, 3) variations in the evolutionary status of the disk, though the trend with inclination is contrary to this possibility, and/or 4) thermal decoupling of the gas and dust such that Tgas>Tdust produces a puffed up gas layer. For the most part, these possibilities are not mutually exclusive and in the next section we explore the role of grain growth and settling.

## 4. Theoretical Modeling

The observational results provide a high-resolution line of sight probe of the structure of circumstellar disks and an initial template for the rate of dust depletion in flared disks. What physical parameters can be inferred from these results? In a truly quiescent disk, all dust particles would settle quickly to a thin layer in the disk's midplane (Miyake & Nakagawa 1995). However, for a disk to accrete at any reasonable rate, the disk must have a non-zero turbulent velocity (Pringle 1981; Hartmann & Kenyon 1996). Suggestions as to the source of the turbulence include the Kelvin Helmholtz instability between the dust and the gas (Wiedenschilling 1997), as well as the MRI instability (Balbus & Hawley 1998). The settling time for the dust at a given radius depends on the level of turbulence and the size of the dust grains. Here we adopt the α-disk formulation in order to evaluate dust settling times, assuming a range of disk alphas given by $\alpha = (4-8) \times 10^{-5}$ and an appropriate disk temperature distribution as a function of radius (D'Alessio et al 1998). Low α values correspond to a system in quiescence. Since the models of dust settling without turbulence indicate dust settling is much too rapid to reproduce the observed gas/dust ratios for disk atmospheres that are at least a few Myr old, we assume quasi-static equilibrium must exist. As the particles settle from the upper atmosphere a balance is reached with diffusion (turbulence) near the midplane.

In the models presented in this section, we start with an MRN power law dust distribution (Mathis, Rumpl & Nordsieck 1977) in quasi steady state, an initial gas/dust ratio equal to the ISM value where the dust is 1% by mass, and the integrated gas and dust surface density is constant. Obviously a strict steady state scenario is violated as dust settling and grain coagulation are expected to be ongoing. But for this restricted epoch, we can use an assumption of quasi-static steady state to imply that the line of sight gas/dust ratio does not change significantly over the time scale of the observed T Tau ages (see discussion in Section 3).

Because the refractory metals (dust) in a protostellar nebula have a mass that is much smaller than the gas mass, the smaller grains can be treated like a passive scalar to a good first approximation. A substantial amount of progress has occurred in our understanding of turbulent mixing of a passive scalar, see recent reviews by Shraiman & Siggia (2000), Warhaft (2000) and Falkovich, Gawedzki & Vergassola (2001). Corresponding advances in our understanding of mixing in astrophysical plasmas have been catalogued in Klessen & Lin (2003) and Balsara & Kim (2005). The same ideas can be used to understand the turbulent diffusion of dust in a turbulent protostellar accretion disk.

In this work, we use the model for dust sedimentation in a turbulent accretion disk given by Dubrulle, Morfill & Sterzik (1995). This model balances the gravitational force acting on the small dust grains against the frictional drag force from the gas, thus obtaining an overall drift velocity for the dust particles towards the midplane of the accretion disk. It also includes the effect of turbulence which diffuses particles in the vertical direction. As a result, grains settle to the midplane with a scale height that is determined by the balance between the downward drift velocity and the turbulence which diffuses particles upwards. Consequently, the downward drift velocity $v_{z,a}$ for a grain of radius "a" in a protostellar accretion disk is given by

$$v_{z,a} = - z \Omega^2 \tau_f \qquad (1)$$

where the characteristic frictional time scale $\tau_f$ is given by:

$$\tau_f = \frac{\rho_s\, a}{\rho_g\, c_s} \text{ in the Epstein regime where } a < \frac{9}{4}\lambda \tag{2}$$

Here $\rho_s$ is the density of the grain particles; $\rho_g$ is the gas density; z is the height above the disk midplane; Ω is the angular rotation velocity in the disk; $c_s$ is the sound speed, and λ is mean free path in the gas. Dubrulle, Morfill & Sterzik (1995) provide a comparison between the scale height $H_a$ for dust of size "a" and H, the gaseous scale height given by:

$$\frac{H_a}{H} = \left(\frac{1}{1+\gamma}\right)^{1/4} \sqrt{\frac{\alpha}{\Omega \tau_f}} \tag{3}$$

Where γ is the spectral index of the turbulence, which we take as the Kolmogorov value 5/3, and α is the disk's viscosity coefficient under a standard α-disk formulation. For this work we assume a minimum mass solar nebula, with a surface density distribution in the disk given by $\Sigma_0 (1AU/r)^{3/2}$, with $\Sigma_0 = 1700$ g cm$^{-2}$ and a temperature distribution taken from D'Alessio et al (1998).

    Thus, the dust scale height is set by a competition between the turbulent diffusivity, which disperses the dust grains above the midplane, and the downward drift velocity. Since the downward drift velocity for bigger grains is much larger, they will have a smaller scale height. We assume that the vertical distribution of the gas varies with "z" as $\rho_0\, e^{-z^2/H^2}$. Following Dubrulle, Morfill & Sterzik (1995), we use a similar functional form for the vertical distribution of the dust so that dust with a size "a" goes as $\rho_{0,a}\, e^{-z^2/H_a^2}$.

    These results enable us to map out the gas distribution in the (r,z) plane as well as make similar maps for dust grains of any size "a". Figure 4a shows the distribution of gas within the inner 300 AU. For illustration, Figures 4b and 4c show the vertical distribution of 0.1μm and 10μm dust grains respectively in the same plane. In Figure 4b it can be seen that the 0.1μm grains track the gas distribution in the inner 200 AU but their scale height becomes smaller than the scale height of the gas at larger distances. At a radius of 10 AU around a 1 M$_\odot$ protostar, the settling time for a fiducial 0.5 micron-sized dust particle is ~0.58 Myr, with larger dust grains having substantially shorter settling times (see also, Miyake and Nakagawa 1995; Youdin and Shu 2002). Consequently, unless all four systems in this study are much younger than 1 Myr, the dust that contributes to the extinction measurements in the extended disks must be considered to be in a quasi-steady state distribution. Figure 4c shows the 10μm grains follow the gas in the inner 15 AU but the scale height becomes much smaller relative to the gas scale height for larger radii. Physically, this stems from the fact that stirring via turbulence does not depend on the grain size while the downward drift velocity, $v_{z,a}$, increases for larger grains.

    The models illustrated in Fig. 4, which include turbulent mixing as well as dust settling, indicate that the expected dust content along lines of sight that are farther from the midplane will be reduced after a few million years. However, to produce a simulated

line of sight dust extinction in the V-band requires an inclusion of grain opacities for dust from 0.1-10 microns. For these grain opacities, we use the results of Draine & Lee (1984) from which we can produce line of sight gas/dust ratios to compare with observations (Table 2).

The theoretical integrated measures of the line of sight gas/dust ratios, normalized to the interstellar value, as a function of inclination are presented in Figure 5a. We present three cases for $\alpha=(4,6,8) \times 10^{-5}$ and an assumed MRN dust distribution. The results, in concordance with observations (Table 2), clearly predict an increasing gas/dust ratio as a function of inclination. It is apparent that the observational data points lie above the models (except for the T Tau N data point versus the $a = 4 \times 10^{-5}$ model) even near the midplane line of sight where turbulent effects are expected to maintain a nearly interstellar gas/dust ratio. Perhaps this indicates that dust settling is not the complete answer and that grain growth needs to be included. Along with dust settling, grain growth amplifies the loss of the smaller particles in the disk atmosphere and as particles grow they settle to the midplane much more rapidly, effectively further decreasing the number of dust particles throughout the disk atmosphere.

The higher density and turbulent conditions in young disks are expected to contribute to rapid particle growth (see Weidenschilling 1997 and references therein). Although the effects of grain coagulation on the residual grain distribution have not been well constrained (see Dullemond & Dominik 2005), a variety of observations have demonstrated the effect of grain growth from the thermal spectral indices and SED's (e.g. Beckwith & Sargent 1991, Manning and Emerson 1994, D'Alessio et al. 2001, Wood et al. 2002), from infrared imaging (McCabe et al. 2003, Duchene et al. 2005), and from mid-infrared spectroscopy (Przygodda et al. 2003) where the shape and strength of the silicate feature in T Tauri stars was used to infer ongoing grain growth.

To incorporate a simple example of grain growth into the model of dust settling, we allow a fraction of the smallest dust particles to coagulate to larger particles. This is accomplished by allowing the initial MRN power law for the dust distribution to decrease from an initial value –3.5 to –3.2, while maintaining a constant total mass (see also D'Alessio et al. 2006). Similar to our range of grain size distributions, Mathis and Whiffen (1989) show that for a power law MRN grain size distribution with an index of -3.7 results in a roughly canonical R ~ 3 extinction law, and for a larger grain size distribution (~2.8) provides a larger R ~ 5. The extinction law with larger average grains and a larger R is consistent with values that are often found to characterize cTTS regions. The decreased slope of the dust distribution reduces the number of 0.1 micron grains by a factor of ~2.7 and effectively increases the gas/dust ratio along all lines of sight since the extinction generated from the dust settling models as well as the $A_V$ measured observationally are dominated by small grains. Grain growth alone primarily affects the normalization of the gas/dust ratio. Infrared spectroscopy and interferometry (Van Boekel et al. 2005) showed silicate particles in disks around HAeBe stars can grow from 0.1 to 2 μm in about $10^6$ years and are dominated by dust sizes of >0.5 μm. This result implies a significant number of the smaller grains have been lost to grain growth in the disk atmosphere. The dust settling line of sight results, with the inclusion of grain growth, are presented in Figure 5b for inclinations from 90 to 20 degrees, indicating a better fit to the four data points than with just an inclusion of dust settling. Because the curvature of the theoretical models is most sensitive to $\alpha$ in the inclination regime of 30-50 degrees, it is apparent (from Fig. 5) that the measured gas/dust ratios in this range will provide the best constraint on the disk $\alpha$,

A more detailed treatment of grain growth is beyond the scope of this paper but the results provide a useful illustration that shows how a realistic model of dust settling and simple model of grain growth can be used to illustrate the observational trend. Obtaining gas/dust ratios for a larger sample of transitional sources will provide better constraints on the turbulent mixing, the need for turbulent transport and radial mixing of dust in the midplane, as well as the need for a more sophisticated model for grain growth. As we noted in Section 3, the observed anomalous gas/extinction ratios are expected to be dominated by the two most obvious issues (grain growth and dust settling) but there are additional, more subtle, ongoing effects that future work will need to address, 1) the consequences of potential age differences among our sources (briefly addressed in Section 3) that may effect the measured gas/dust ratios, 2) differences in stellar masses that may lead to differences in the evolutionary timescales of the gas and dust envelopes, and 3) the decoupling of the gas and dust temperature in the atmosphere of the disk (see Kamp and Dullemond 2004, Glassgold et al. 2004). In the case where $T_{gas}>T_{dust}$, the heightened scale for the gas distribution relative to the dust would affect the gas/dust ratio. The magnitude of these effects will be incorporated in a later paper. Thus, at this stage, the theoretical results are only influenced by dust settling and grain growth in disks around young stars. The effects grain growth and dust settling are not mutually exclusive and indeed grain growth likely leads to dust settling.

**5. Conclusion**

The results presented in this work provide observational evidence and initial quantification of the stratification of gas and dust in disks around stars that are likely in the early stages of planet formation. The implications of this analysis are that the role of dust settling in the extended disks can be constrained by the observations and that models combining disk turbulence with dust settling and grain growth can reasonably account for the observed gas/dust ratios. An understanding of how the dust settles to the midplane relative to the gas has the potential to provide essential constraints for physical parameters in the midplane, specifically, the results emphasize the potential power of this technique to place observationally well-motivated bounds on the accretion disk alpha. Additional observations of the gas/dust ratio from a larger sample of young disks will enable us to more fully constrain turbulent mixing, the grain mass distribution, the rate of grain coagulation and the mechanism for planetesimal formation. We are well aware of the obvious limitations of the inclination and extinction measures in context with only four observations and encourage future work to increase the statistics.

Future modeling efforts will incorporate a more detailed dynamical modeling of turbulence, and a self-consistent turbulent dispersal of dust grains of various sizes. With a more statistically significant quantification of the gas/dust ratio as a function of height above the midplane, further constraints can be placed on the development of model SEDs (Chiang et al. 2001) as well as provide important implications for models of the photochemistry of disk atmospheres and their temperatures that generally assume gas and dust mixing (Kamp & Dullemond 2004; Gorti & Hollenbach 2004; Glassgold et al. 2004).

**Acknowledgements**: The data presented herein were obtained at the W.M. Keck Observatory, which is operated as a scientific partnership among the California Institute of Technology, the University of California and the National Aeronautics and Space Administration. The Observatory was made possible by the generous financial support of the W.M. Keck Foundation. DSB acknowledges support via NSF grants R36643-7390002, AST-005569-001, DMS-0204640 and NSF-PFC grant PHY02-16783. TWR, SDB and ELG were supported in part by an NSF Astronomy grant AST02-05581. SB was also supported in part by a NASA Michelson Fellowship (2005-2006). We also want to thank the anonymous reviewer for many helpful comments that improved the theory section greatly. We also wish to thank Andrew Youdin and Joan Najita for their comments and suggestions.

FIGURE CAPTIONS:

**Figure 1. a)** M-band spectra showing the fundamental $^{12}$CO absorption lines from the four sources. The spectra have been normalized and shifted vertically for clarity. RNO 91 shows an offset (blue dotted line) that is due to a difference in observed radial velocity relative to the other sources (see Table 1). The red dotted lines indicate the centered line positions for the CO lines. The broad depression seen in RNO 91 centered on 2140 cm$^{-1}$ is due to CO ice. **b)** K-band spectra showing the $^{12}$CO overtone v=2-0 absorption lines from the four sources. The spectra have been normalized and shifted for clarity. The dotted lines show the velocity offset for RNO 91, which is due to a difference in radial velocity at the time of observation.

**Figure 2.** Example of a cross-trace through one NIRSPEC spectral order of T Tau N and T Tau S. The slit was fixed N-S for these observations of T Tau. The solid histogram is the data, the dotted lines are Gaussian best fits to the double peaks separated by 0.7 arc seconds. The dash-dotted line shows the resulting the pixel match. The rows extracted for the analysis were 31-36 and rows 42-49 so contamination was minimal.

**Figure 3.** Line-of-sight to each source.

**Figure 4. a)** The log of the gas number density in the (r,z) plane within the inner 300 AU. **b)** The log of the dust number density for the 0.1 μm dust grains and $\alpha = 6 \times 10^{-5}$ in the same (r,z) plane within the inner 300 AU. The dust scale height for the small grains is bounded from above by the gas scale height because grains that are well-mixed with the gas should then track the scale height of the gas that bears them aloft. The systems being observed have lifetimes ~3-5 million years, much longer than settling times (~0.5 Myr) for the grain sizes of interest. As a result, since we see gas and dust at all the observed inclinations, we can assume a quasi steady state distribution for the dust grains, ie., a balance between downward drift of grains and turbulence. **c)** The log of the dust number density for the 10μm dust grains and $\alpha = 6 \times 10^{-5}$ in the same (r,z) plane within the inner 300 AU.

**Figure 5** The variation of the theoretical normalized gas to dust ratio, $\Delta$, as a function of the inclination angle of the disk with the line of sight for $\alpha=(4,6,8) \times 10^{-5}$. The observational data points, similarly normalized, are also provided. The $A_V$ we use originates from the same particle distribution that is most affected by dust settling in the model and is also the range of dust sizes that contribute the maximum number of dust particles in an MRN distribution. 5b) The effect of adding grain growth to the models compared to the observational data. The initial MRN power law for the dust distribution is allowed to decrease from an initial value –3.5 to –3.2 to simulate a reduction in the number of smaller grains as expected with grain growth.

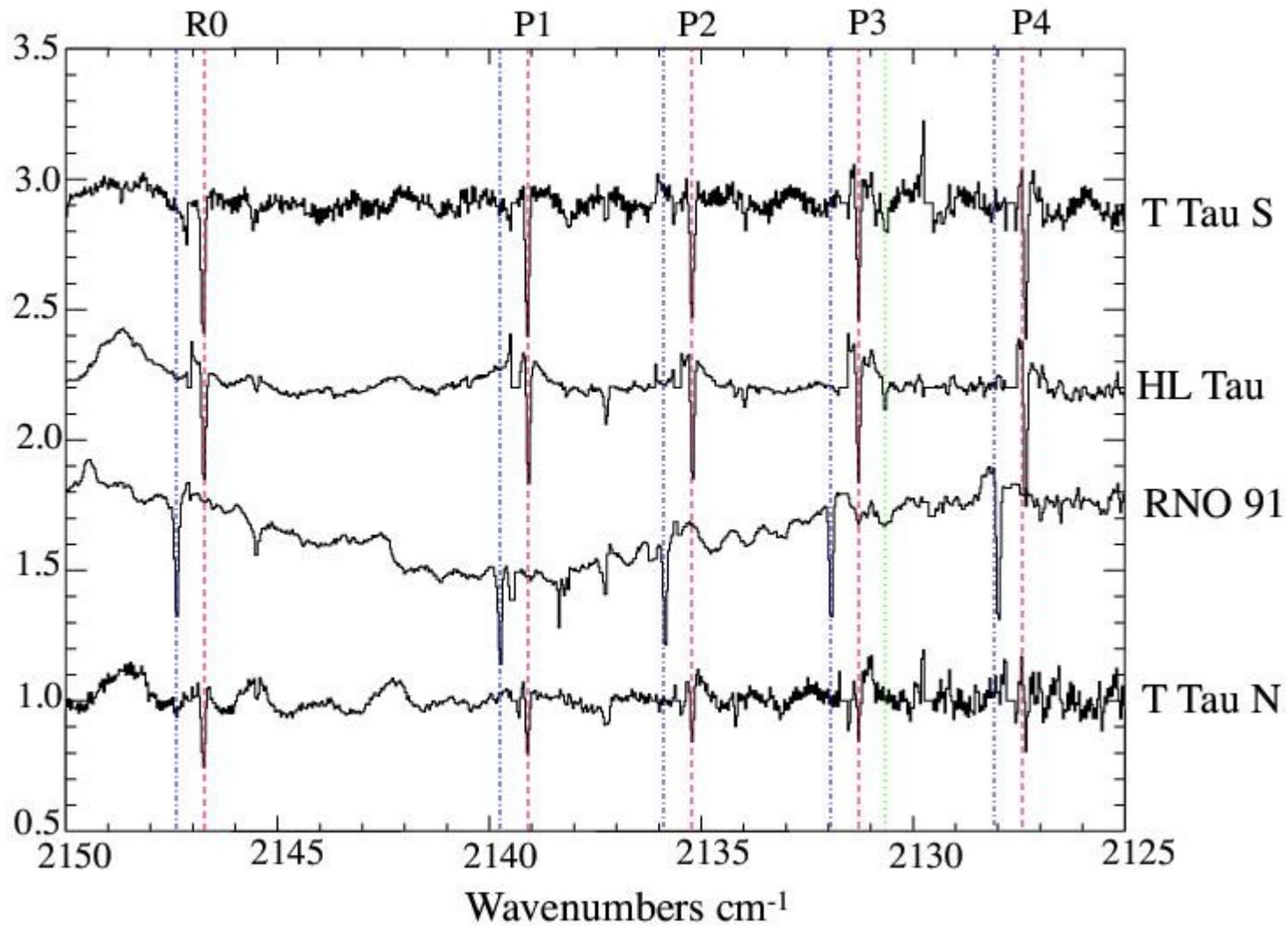

Fig. 1a

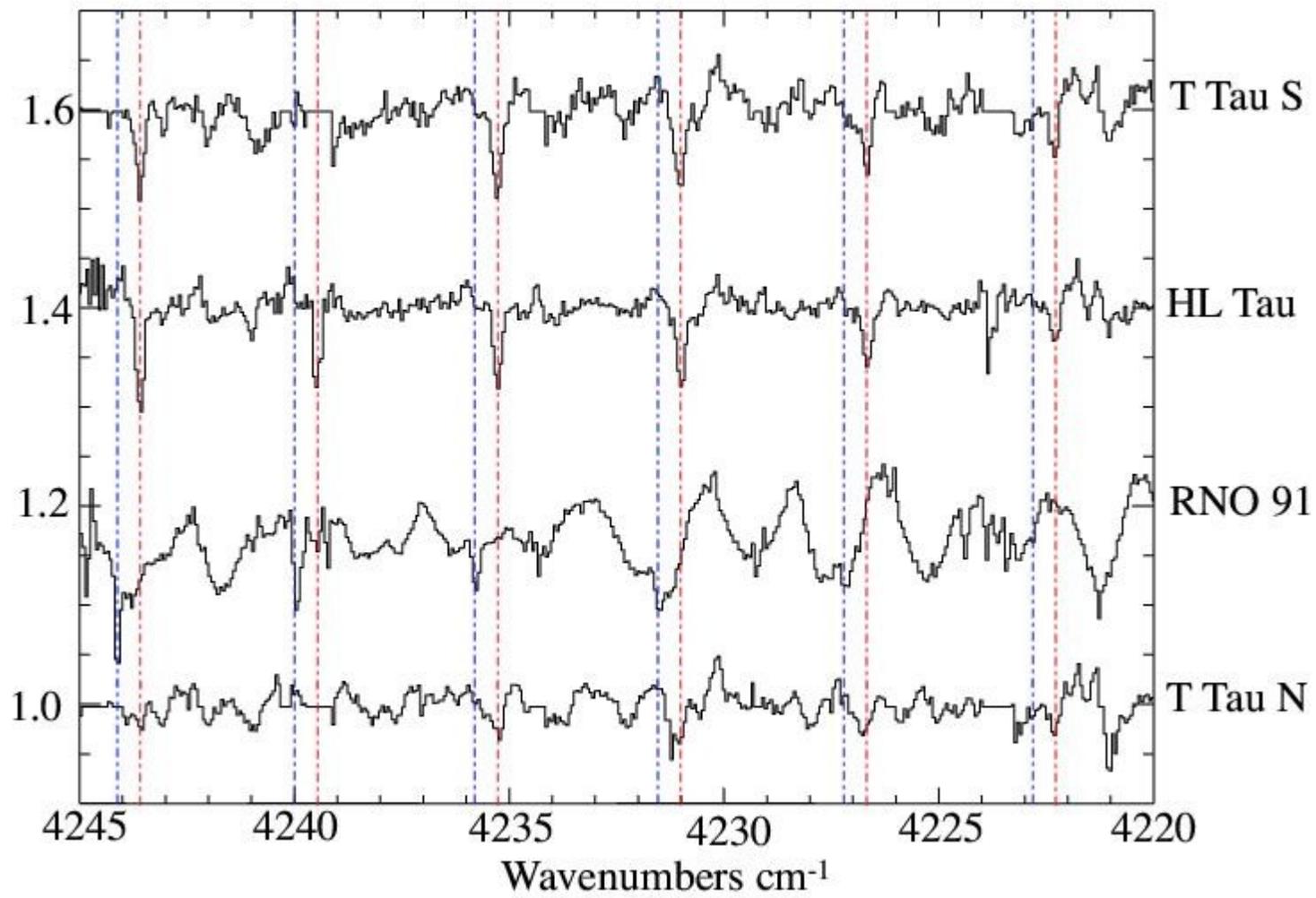

Fig. 1b

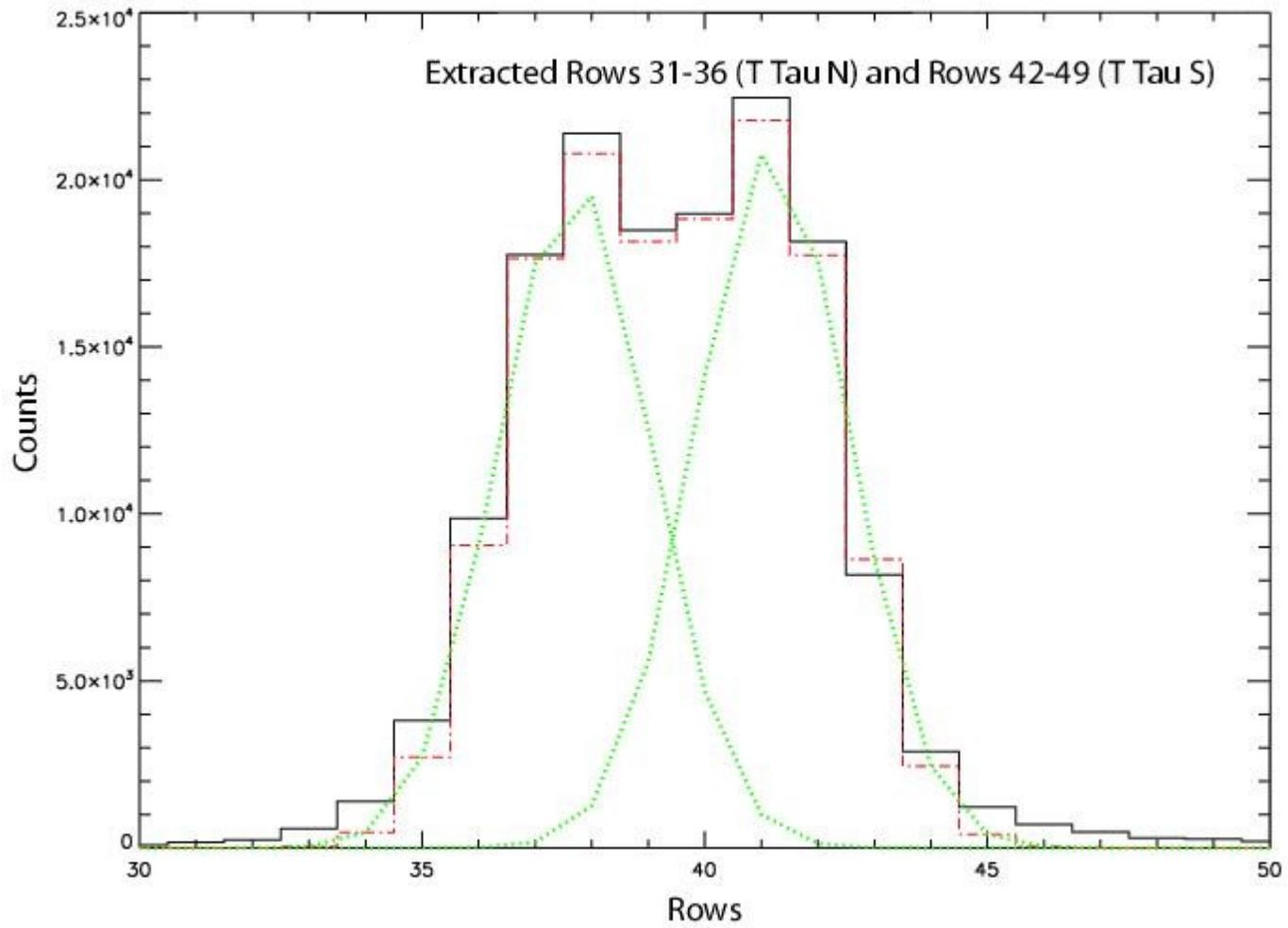

Fig. 2

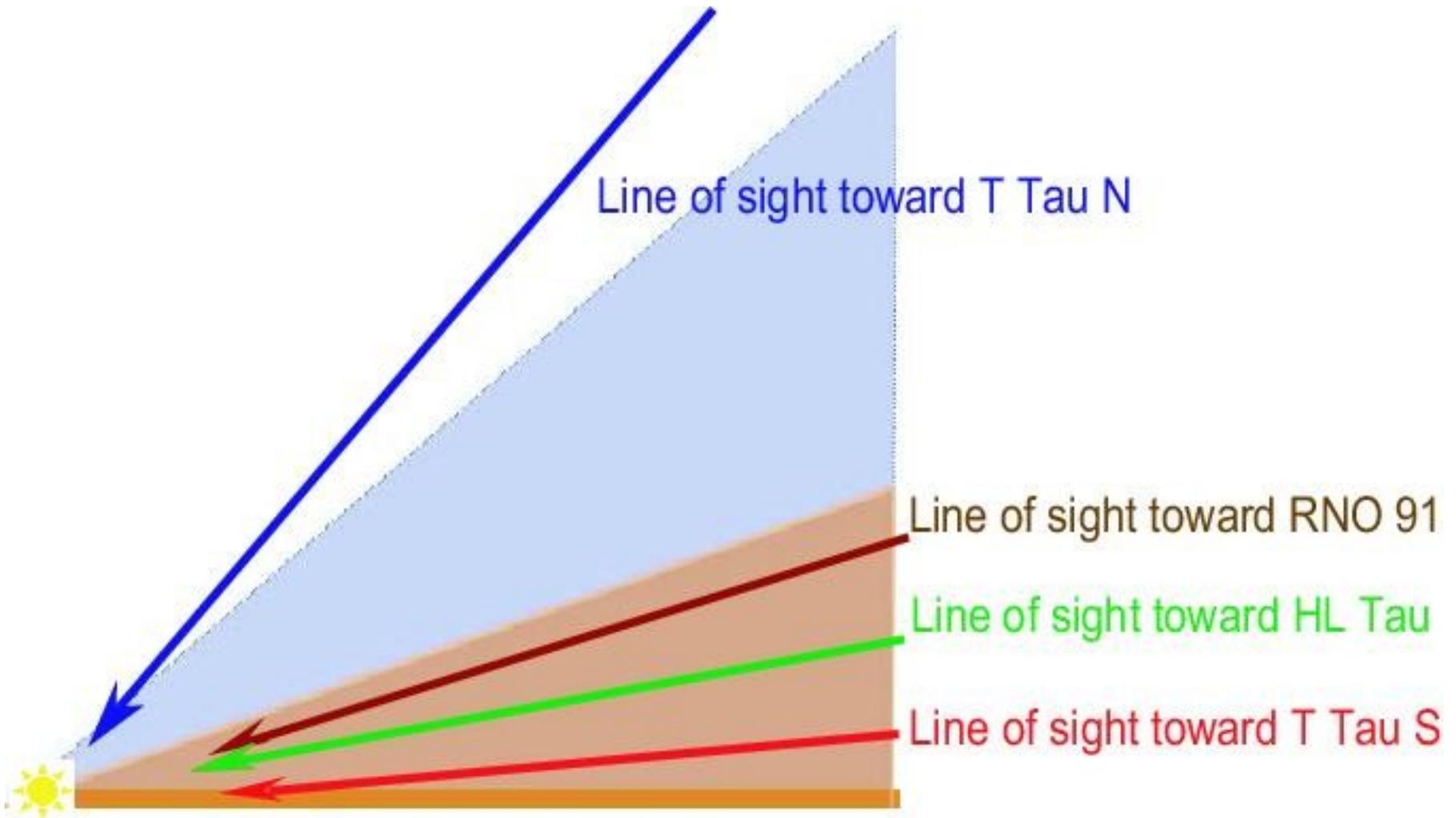

Fig. 3

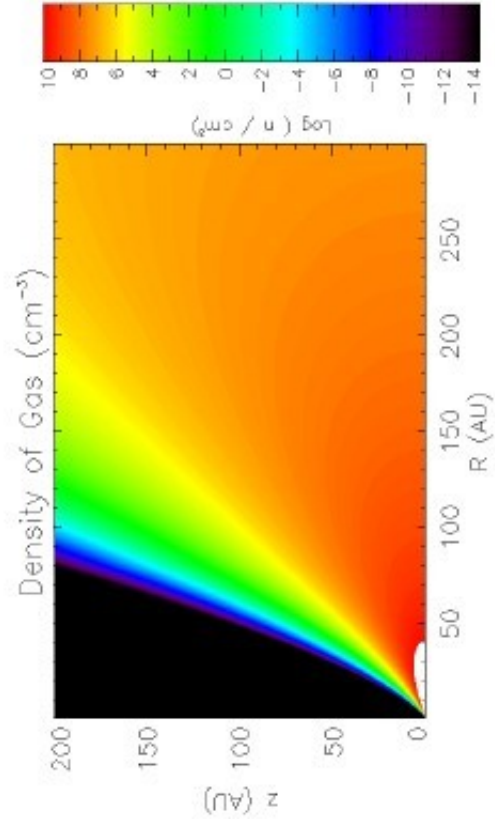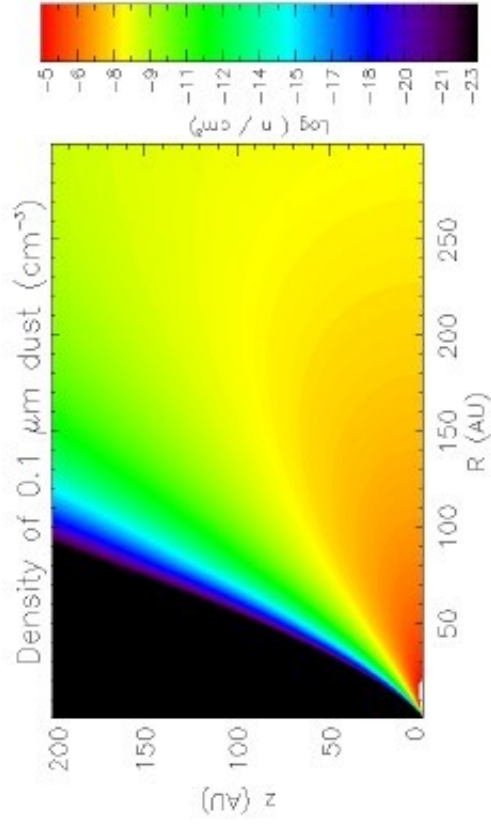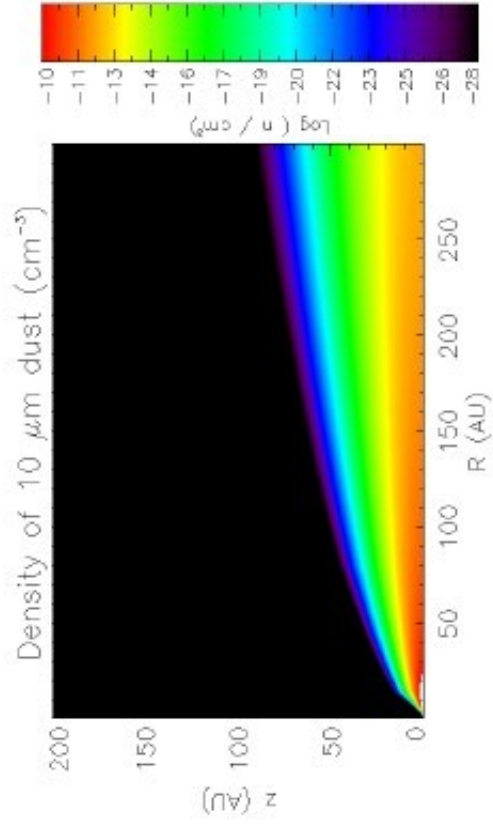

Fig. 4

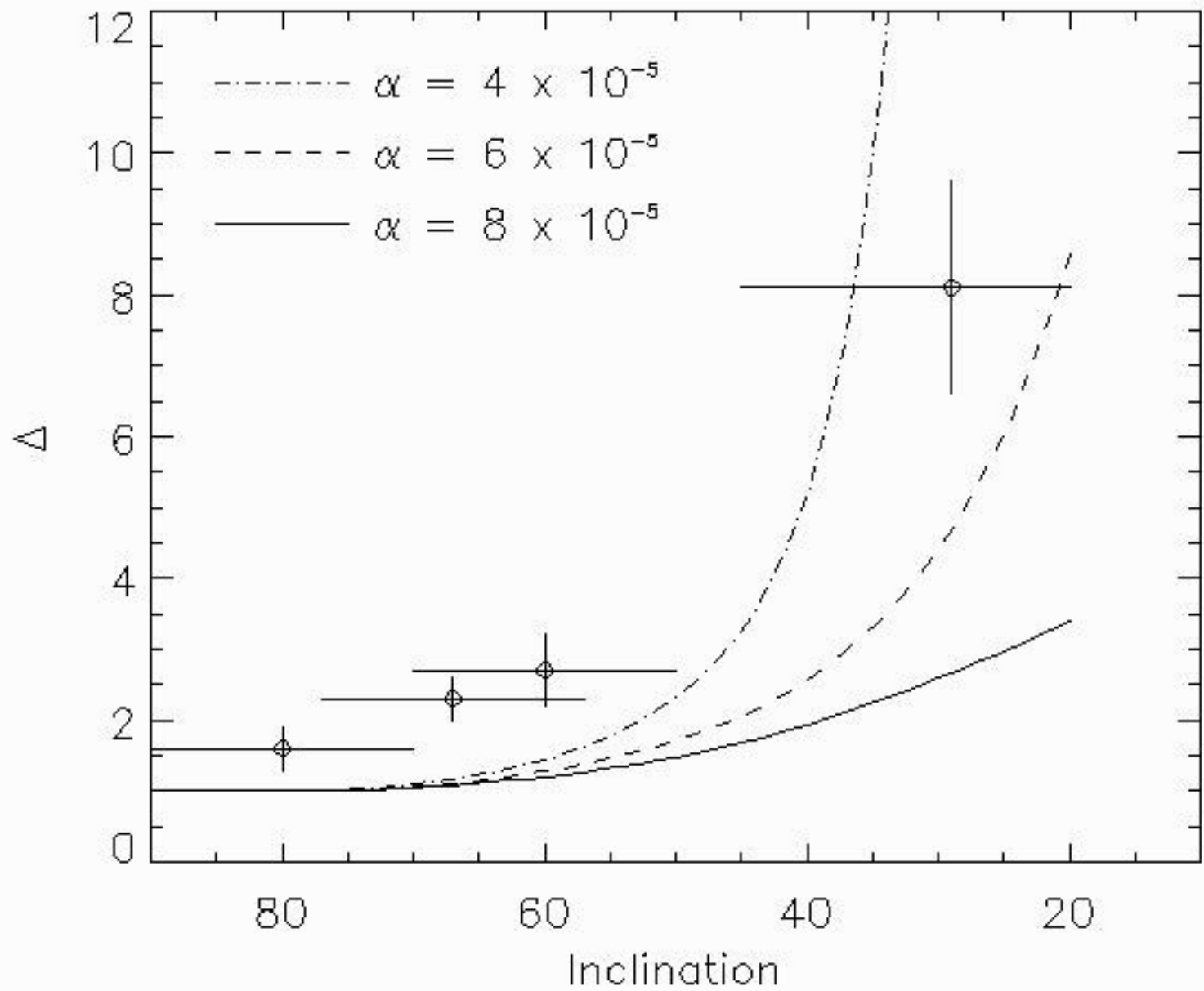

Fig. 5a

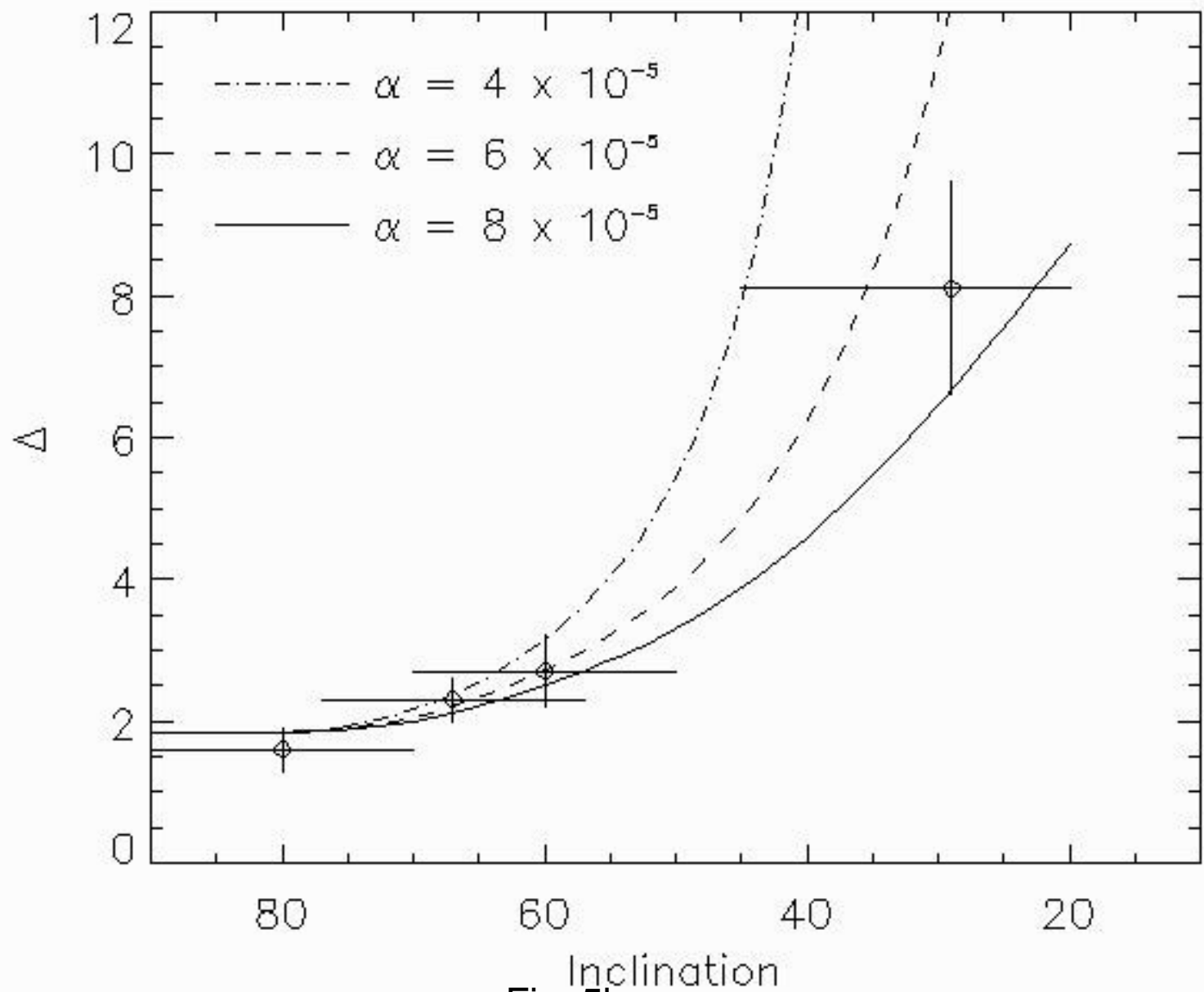

Fig. 5b